\documentclass[ journal ]{IEEEtran}

\usepackage{graphicx,cite,epsfig,amssymb,amsmath,multicol,subfigure,mathtools,bm,mathrsfs,multirow}
\usepackage{algorithm}
\usepackage{algorithmic}
\usepackage{verbatim}
\usepackage{xcolor}
\allowdisplaybreaks[4]

\makeatletter
\renewcommand{\maketag@@@}[1]{\hbox{\m@th\normalsize\normalfont#1}}%
\makeatother

\begin{document}


\title{\huge Challenging the Law of Energy Conservation Through Superposed Waves Based on Spatial Symmetry of Two RF Sources: Theoretical Derivation and Experimental Verification}

%



\author{Bingli Jiao, Chenbo Wang, Zijian Zhou
	\thanks{B. Jiao, C. Wang and Z. Zhou are with the Department of Electronics, Peking University, Beijing 100871, China (e-mail: jiaobl@pku.edu.cn, wcb15@pku.edu.cn, zjzhou1008@pku.edu.cn).} 
}			

\maketitle

\begin{abstract}
This study is grounded in the concept of spatial symmetry, which allows two co-phase RF sources to jointly radiate harmonic electromagnetic (EM) waves, even in presence of electromagnetic couplings between them.  The superposition law is directly applied to the two waves owing to their sources, including the EM coupling effects.  Our research uncovers a conflict with energy conservation law at the following two levels: (1) the total radiative powers as defined by Poynting theorem, and (2) further, the input powers of the sources. To explore this phenomenon, we create a symmetric dipole model to examine the energy behaviors at the sources and the superposed waves, separately.  Both of the results reveal that the energy-doubling phenomenon presents when the dipoles are placed in close proximity. As the distance between the two dipoles increases, it is found that the total radiative power fluctuates in a damped manner and ultimately converges to a value required by energy conservation law. The theoretical conclusion of energy doubling is validated by experimental observations, which show a 1.59-fold increase in power within a microwave anechoic chamber. By analyzing the interactions, i.e., the EM coupling, between the two sources, we characterize a complete phenomenon of energy non-conservation. 

\end{abstract}


\section{Introduction}
\label{introduction}
In a linear system, it is well known that the superposition of two waves having identical amplitudes and frequencies can result in a superposed wave whose local power density is quadrupled somewhere due to the co-phase interference, in comparison with that of a single wave. Conversely, with  $180^o$ anti-phase interference, the power density can be completely nullified. This suggests that energy conservation does not necessarily hold locally. While, this paper goes further to explore the energy conservation issue over the entire space. For simplicity,  we restrict our discussions solely on the constructive interference, since the deconstruction problem can be treated using the same method of this paper. In addition, the near fields are not discussed, as their powers are confined locally.

In 1980, Levine discussed a superposition phenomenon in the radiation model, where two harmonic waves undergo constructive interference throughout space by placing the two sources very close to each other, with a distance much smaller than the wavelength, i.e., $d/\lambda <<1 $, where $d$ is the distance and $\lambda$ is the wavelength [1].  In fact, as the two waves' amplitudes are linearly superposed, an amplitude-doubling effect results, causing the total radiative power to increase by approximately four times that of one single wave, rather than two times required by energy conservation law.  Therefore, the ``energy doubling" problem arises because the energy-doubling effect has extended to whole space instead of some local areas.

Wishfully attempting to comply with the law of energy conservation, the energy doubling phenomenon was dismissed as a ``false paradox"  due to an incorrect explanation using the interactions between the two sources[1], as outlined below. Firstly, the explanation separates the EM coupled sources from the application of the superposition law, leading to a power estimation that does not reflect true powers of the sources. Secondly, the two antennas are arranged parallel to one another, maximizing EM coupling between them for getting its weight.  These two points mixed together resulted in the vague and incorrect explanation.  In fact, the error becomes clear when the two whip antennas are placed in a straight line. Since each antenna’s radiation is zero in the direction of its pole—i.e., toward its neighboring antenna—there is no any EM interactions between them. Nevertheless, the energy-doubling effect still persists, as will be demonstrated later.

Due to the incorrect explanation in [1], some studies have been misled, resulting in the failure to achieve either analytical solutions [2-4] or experimental verification. Meanwhile, other researchers continue to debate the issue in their papers, offering viewpoints that are either inconsistent with or contradictory to the law of energy conservation [5-8]. In such, this issue can still be considered an open problem within the realm of classical electromagnetism.

As is widely acknowledged, the laws of superposition and energy conservation operate independently of each other. In other words, their validity and application do not necessarily promise one another. This realization drives us to conduct in - depth research on the former and utilize the results to verify the latter, even though that this approach is boldly challenging the fundamental principles of physics.

To apply the superposition law correctly, we must apply it directly to the two waves radiated from their sources including the coupling effects between them.  In the presence of (EM) coupling, the sources can influence each other. Although the input power of each source may differ from the case where each source radiates independently, the power-consumption of each source should equal the power it radiates. In contrast, if there is no EM coupling, each source radiates as if it were alone, resulting in neither increase nor decrease in power consumption compared to the scenario where a single source radiates on its own.  In both cases, \textbf{the total power-consumption of the two sources should equal the power transferred to the radiative waves, regardless of weather any EM coupling exists}.  

By summarizing the analysis above, we can clearly conclude that if the power summation of two radiatve waves can be doubled through the superposing and, then, the input powers of the sources can be regarded to be doubled as well.  When the evidences of energy-doubling are sufficient, we should believe the fact, rather than adhere with the law of energy conservation tightly. This bold move avoids being entangled in the complexities of EM coupling and makes our derivations straightforward in achieving the results, introducing new perspectives into classical electromagnetism [9]–[11].

To be specific, we proposed a spatially symmetric system, in which a dipole model is constructed to ensure the correct application of the superposition law, incorporating the following key characteristics.

\begin{itemize}
	
	\item Upon spatial symmetry, two identical harmonic dipoles can exist with co-phase radiations physically.  
	
	\item The antennas' design minimizes the EM coupling for simplifying experimental data.
	
	\item In assumption of small distance between the dipoles, the total radiative power can be easily calculated at the sources.
	
	\item More results are obtained by increasing the distance between the two dipoles.   
	
	\item The experiments taken in a microwave anechoic chamber substantiate the theoretical conclusion.
	
\end{itemize}

Finally, some key technical terms and specifications of this paper are listed in Table I, and the detailed studies are presented in the following sections. 

\begin{table}[h!]
	\centering
	\caption{Abbreviations and SI Units}
	\label{table:abbreviations}
	\begin{tabular}{|c|c|}
		\hline
		\textbf{Abbreviation} & \textbf{Description} \\
		\hline
		EM & Electromagnetic \\
		EAF & Energy Abnormal Factor \\
		GHz & Gigahertz (1 GHz = $10^9$ Hz) \\
		m & Meter \\
		°C & Degree Celsius \\
		dB & Decibel \\
		$\Omega$ & Ohm \\
		\hline
	\end{tabular}
\end{table}


\section{Energy-Doubling Phenomenon}
\label{Spatial Symmetry}
In contrast to conventional studies, spatial symmetry plays a conceptual role as the physics basis that allows two identical sources to radiate in co-phase manner,  in the presence of EM coupling between them.  Hence, we can build the theoretical study directly on the two existing identical sources. The following discussions are confined to the linear approximation of radiation in free space.

A dipole model is set up by positioning two co-phase dipoles along the z-axis, with the symmetry point located at the origin, as shown in Fig. 1, where the closed spherical surface, $\mathfrak{S}$, bounds the two dipoles. 

\begin{figure}[!t]
	\centering
	\includegraphics[width=0.35\linewidth]{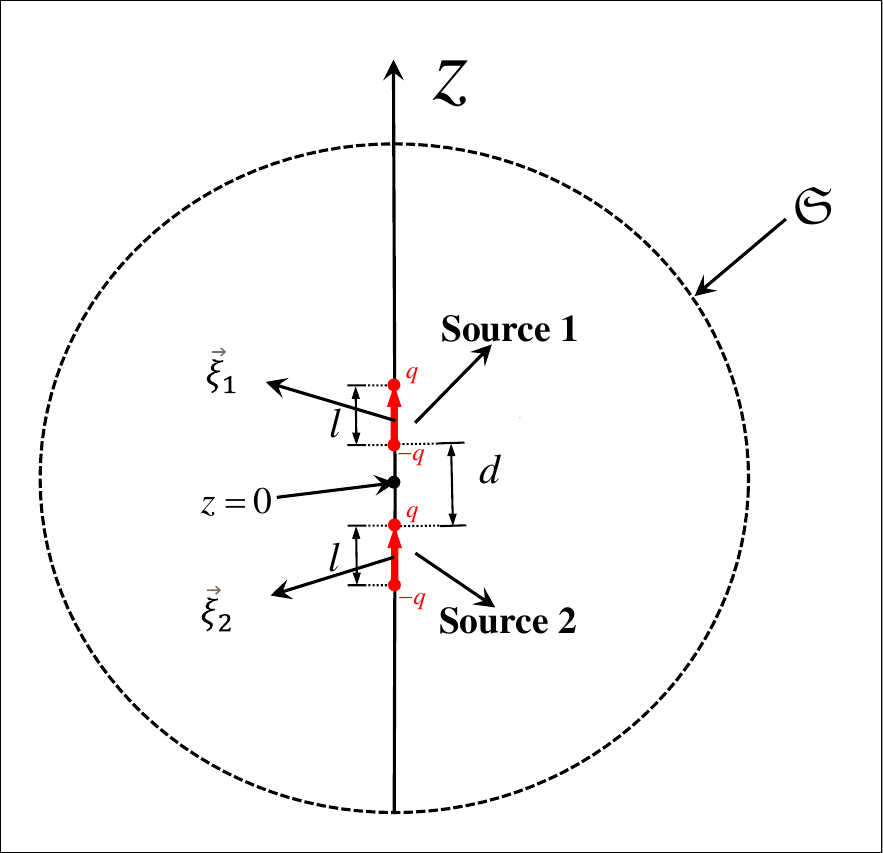}
	\caption{The dipole model with spatial symmetry at $z=0$}
	\label{fig_radiation_1}
\end{figure}  

At first, let us focus on the sources, i.e., the two dipoles, by assuming
\begin{equation}\label{asum}
	d \ne 0,  \ \ \ d<< l  \ \ \ \text{and}  \ \ d/\lambda<<1
\end{equation}
where $d$, $\lambda$ and $l$ are the distance between the two dipoles, the wavelength and the distance between the two charges of each dipole, respectively. 

As long as spatial symmetry applies, each dipole radiates the same power, given by [11]
\begin{equation}\label{di-1}
	\centering
	P_1 = P_2 = \frac{(ql)^2 \mu_0\omega^4}{ 12\pi C}    \ \ \   
\end{equation}
where $P_1$ and $P_2$ denote the radiative powers of dipole 1 and dipole 2, $C$ and $\mu_0$ are the velocity of light and the permeability, $q$ and $l$ are the magnitude of each electric charge and the distance between the two charges of each dipole, and $\omega$ is the angular frequency, respectively. 

When $d/\lambda \to 0$ and $d/l \to 0 $, the two dipoles approximately coalesce into an equivalent dipole in form of $q (2l) $. This occurs because the two charges, positioned at minimal distance, nearly cancel each other due to opposite signs. Meanwhile, the separation distance between the remaining charges approaches $2l$ (see Fig.1),  The total radiative power can thus be approximated by
\begin{equation}\label{di-2}
	P \to \frac{ [q(2l)]^2 \mu_0\omega^4}{ 12\pi C}  = 4\frac{ (ql)^2 \mu_0\omega^4}{ 12\pi C} = 4P_i \    \text{for}   \ i \ = \ 1, 2
\end{equation} 
where $P$ is the radiative power of the equivalent dipole. 

Equation (3) reveals surprisingly the energy-doubling phenomenon, as the equivalent dipole takes the role of the radiations of the two dipoles. However, its radiative power is four times that of a single dipole, rather than the two times required by the energy conservation law.

Next, we examine the radiative fields upon the superposition law, where Poynting theorem places a crucial role relating the power of each dipole to its raidative wave.  Upon spatial symmetry, the EM excitations of the two dipoles should be identical (see Fig. 1), as expressed by
\begin{equation}
	\begin{split}\label{01}
		\overrightarrow{E} (\overrightarrow{\xi}_1,t) =   \overrightarrow{E}(\overrightarrow{\xi}_2,t)  
	\end{split}
\end{equation}
\begin{equation}
	\begin{split}\label{02}
		\overrightarrow{J}(\overrightarrow{\xi}_1,t) =   \overrightarrow{J}(\overrightarrow{\xi}_2,t) ,
	\end{split}
\end{equation}     
in space- and time domain, where $\overrightarrow{E} (\overrightarrow{\xi}_i,t)$ and $\overrightarrow{J}(\overrightarrow{\xi}_i,t)$ represent the electric field strength and current density at the geometric center,  $\overrightarrow{\xi}_i$,  of dipole
$i$, for $i=1,2$, respectively.

Applying the superposition law to the two waves radiated from the two dipoles yields 
\begin{equation}
	\begin{split}\label{03-1}
		\overrightarrow{E} = \overrightarrow{E}_1 +\overrightarrow{E}_2 
	\end{split}
\end{equation}   
and
\begin{equation}
	\begin{split}\label{03-2}
		\overrightarrow{H} = \overrightarrow{H}_1 +\overrightarrow{H}_2 
	\end{split}
\end{equation}    
where $\overrightarrow{E}$ and $\overrightarrow{H}$ are the resultant radiative electric- and magnetic filed in the free space, and $\overrightarrow{E}_i$ and $\overrightarrow{H}_i$ are the electric- and magnetic filed owing to the radiation from the dipole centered at $\overrightarrow{\xi}_i$, for $i=1,2$.   

Apparently, the energy conservation law may require the superposed wave to satisfy Poynting’s integral by [11]
\begin{equation}
	\label{041}
	\mathop{{\int\!\!\!\!\!\int}\mkern-21mu\bigcirc}\limits_\mathfrak{S}
	\left\langle \overrightarrow{S}\right\rangle \cdot d\overrightarrow{s}+\iiint\limits_{\xi'_1}{{}} \left\langle \overrightarrow{J}\cdot \overrightarrow{E}\right\rangle dv + \iiint\limits_{\xi'_2}{{}} \left\langle \overrightarrow{J}\cdot \overrightarrow{E}\right\rangle dv= 0 \ \ \ \ \ 
\end{equation}
where $\left\langle\cdot \right\rangle$ is the operator taking the average value, $\overrightarrow{S} = \overrightarrow{E}\times \overrightarrow{H}$ represents the Poynting's vector of the superposed wave and $\xi'_i$ the coordinates within dipole $i$ centered at $\xi_i$, respectively.

The first term on the right side of equation (8) represents the total radiative power, while the second and third terms presents the power-consumption of dipole $1$ and $2$, respectively.

Substituting equations (6) and (7) into the first term on the right side of (8) yields
\begin{equation}
	\label{041}
	\mathop{{\int\!\!\!\!\!\int}\mkern-21mu\bigcirc}\limits_\mathfrak{S}
	\left\langle \overrightarrow{S}\right\rangle \cdot d\overrightarrow{s}= \mathop{{\int\!\!\!\!\!\int}\mkern-21mu\bigcirc}\limits_\mathfrak{S}
	\left\langle \overrightarrow{S}_1\right\rangle \cdot d\overrightarrow{s}+	\mathop{{\int\!\!\!\!\!\int}\mkern-21mu\bigcirc}\limits_\mathfrak{S}
	\left\langle \overrightarrow{S}_2\right\rangle \cdot d\overrightarrow{s}+\mathfrak{M}_{12}\ \ \ \ \ 
\end{equation}
with $\overrightarrow{S}_i =\overrightarrow{E}_i\times \overrightarrow{H}_i$ and $ \mathfrak{M}_{12} =
\mathop{{\iint}\mkern-17mu\bigcirc}\limits_\mathfrak{S}
\left\langle {\overrightarrow{E}_1 \times \overrightarrow{H}_2 + \overrightarrow{E}_2 \times \overrightarrow{H}_1}\right\rangle \cdot d\overrightarrow{s}$, where  $\mathfrak{M}_{12}$ is the result of the interference between the two radiative waves. 

Based on Poynting theorem, the radiative power of each dipole relates to its power-consumption by 
\begin{equation}
	\begin{split}\label{043}
		P_i = \mathop{{\int\!\!\!\!\!\int}\mkern-21mu\bigcirc}\limits_\mathfrak{S}
		\left\langle \overrightarrow{S}_i\right\rangle \cdot d\overrightarrow{s}   = -\iiint\limits_{\xi'_i}{{}} \left\langle \overrightarrow{J}\cdot \overrightarrow{E}\right\rangle dv  \ \ \
	\end{split}
\end{equation}
where $P_i$ is the radiative power of dipole $i$.

Considering (8) to (10), we can find that the law of energy conservation stated in (8) requires
\begin{equation}
	\mathfrak{M}_{12} \equiv 0
\end{equation}
which can be conflicted when the distance between the two dipoles is very small, i.e., $d/\lambda <<1$, because of 
\begin{equation}
	\begin{split}\label{01}
		\overrightarrow{E}_1 \approx \overrightarrow{E}_2  \ \ \text{and} \ \ \overrightarrow{H}_1 \approx \overrightarrow{H}_2  
	\end{split}
\end{equation}
everywhere in whole space, leading to 
\begin{equation}
	\mathfrak{M}_{12} \approx 2 P_i
\end{equation}
for $i=1$ or $2$. 

By defining the total rediative power as
\begin{equation}
	\begin{split}\label{042}
		P_T = \mathop{{\int\!\!\!\!\!\int}\mkern-21mu\bigcirc}\limits_\mathfrak{S}
		\left\langle \overrightarrow{S}\right\rangle \cdot d\overrightarrow{s}, 
	\end{split}
\end{equation}
where $P_T$ is the total radiation power of the superposed wave calculated in (9), we found  
\begin{equation}
	\begin{split}\label{42}
		P_T = 4P_i ,
	\end{split}
\end{equation}
which indicates, again, the total radiative power of the superposed wave is 4 times that from the individual dipole.

The theoretical conclusions derived from both the sources and the radiative fields agree each other exactly, demonstrating the energy-doubling phenomenon, which is the fact contradicting the energy conservation law.

\section{Energy Abnormal and Conservation}
In this section, we further explore the effect of increasing the distance between the two sources in the symmetric structure. The dipoles devised in the previous section are replaced with two antennas, each with $0.8$ cm in length, and excited by a frequency of $2.4$ GHz. 

By employing the simulation method to generate the EM waves radiated from these two antennas, we obtain the numerical results of the total radiative power of the superposed waves by using Poynting integrals. 

To be Specific, we define an energy abnormal region where the law of energy conservation is not necessarily valid, and introduce an energy abnormal factor (EAF) to quantify the phenomena by
\begin{equation}\label{dipole-2}
	\beta=\frac{P_T}{P_{con}} = \frac{P_T}{P_1+P_2} \ \ \ 
\end{equation}
where $\beta$ is the EAF, $P_T$ is the actual total radiative power of the superposed wave, $P_{con}$ the power-value required by energy conservation law  and $P_i$ denotes the radiative power of antenna $i$.  

Using the simulated radiative waves, the total radiative power of the superposed wave is calculated by substituting equation (9) to (14), while the radiative power of each individual wave is computed using the first equality on left side of (10). From these results, the values of $\beta$ are obtained using equation (16) for various values of $d/\lambda$.

Figure 2 shows the numerical results of (16),  which indicates that \textbf{the total radiative power $P_T$ fluctuates} with increasing $d/\lambda$, alternating between $\beta > 1 $ and $\beta < 1 $, and \textbf{converging eventually to $\beta =1 $ required by the law of energy conservation}.  The region from $d/\lambda \approx 0$ to $d/\lambda = 3$ can be defined as the energy abnormal region, because the fluctuation of $\beta$ is obviously found.  While, the energy conservation law dominates in the region from $ d/\lambda>3$ to $ d/\lambda = \infty$, as values of $\beta$ are very close to 1 and continuously approaching to $1$.     

The energy behavior in the abnormal region is difficult to understand, especially at $d/\lambda =1 $, where the two radiation resources can reinforce each other slightly, yet the total rediative power is smaller than the value required by the energy conservation.  This paper will not present further discussions on the abnormal region due to lack of detailed studies.   

To be further convinced, it is found that the numerical in Fig.2 are essentially the same as those of [12], which, unfortunately, reached an incorrect conclusion due to not consider the spatial symmetry.

\begin{figure}[htbp] 
	\centering
	\includegraphics[width=0.35\linewidth]{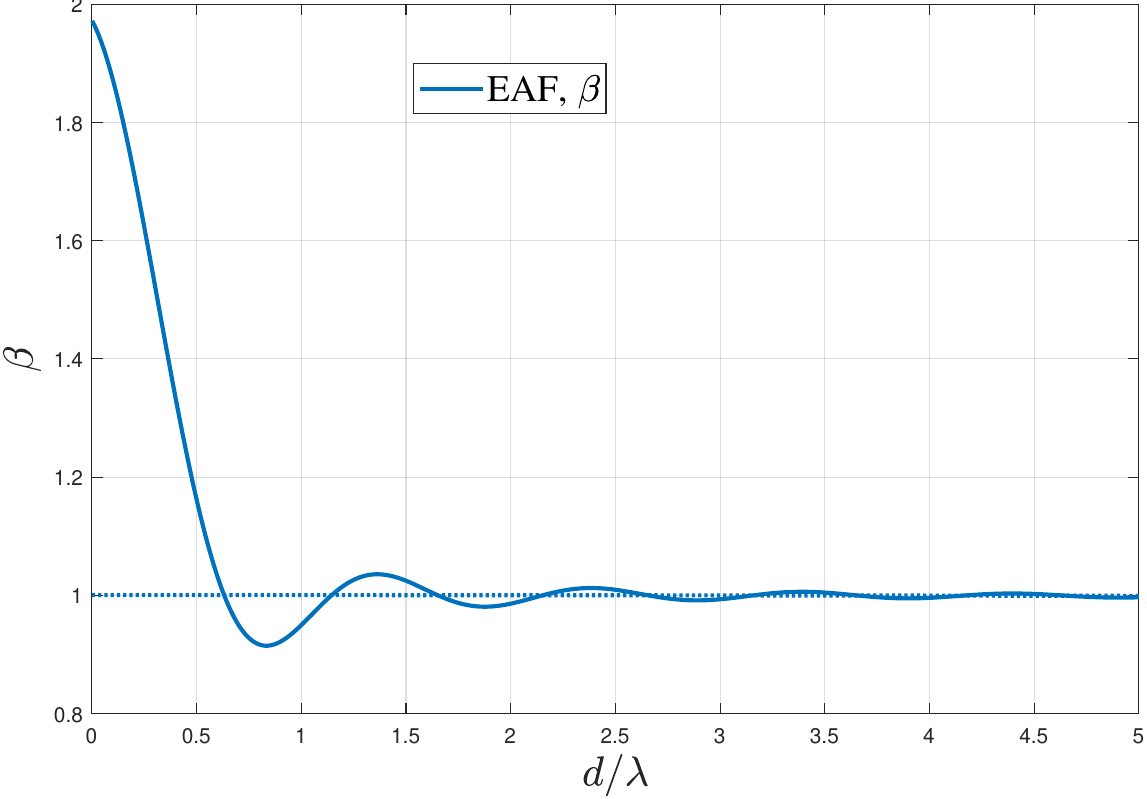}
	\caption{The numerical results of $\beta$ vs. $d/\lambda$ for showing the fluctuation of the total radiative power.}
	\label{fig_power}
\end{figure}

\section{Experiment Verification}
\begin{figure}[htbp] 
	\centering
	\includegraphics[width=0.45\linewidth]{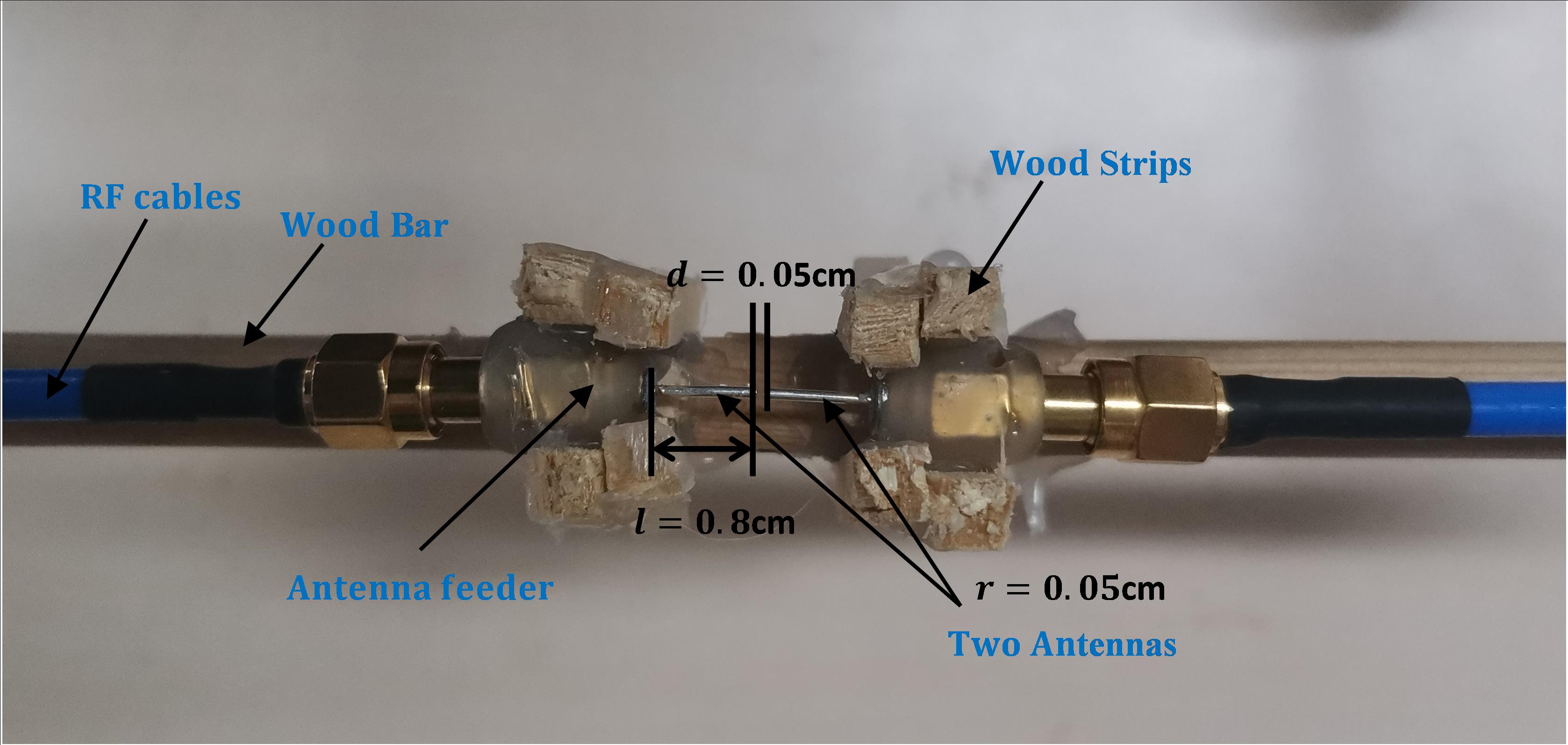}
	\caption{A Photo of the two antennas.}
	\label{fig_3}
\end{figure}

In this section, we conduct the experiment to verify the theoretical conclusion of the energy-doubling effect. Two thin copper wires, each 0.8 cm in length, are excited by a frequency of $2.4$ GHz, the same as assumed at the simulations in the last section. Figure \ref{fig_3} shows a photo of the two antennas with a separation distance of $d=0.05$cm. 

\begin{figure}[htbp] 
	\centering
	\includegraphics[width=0.45\linewidth]{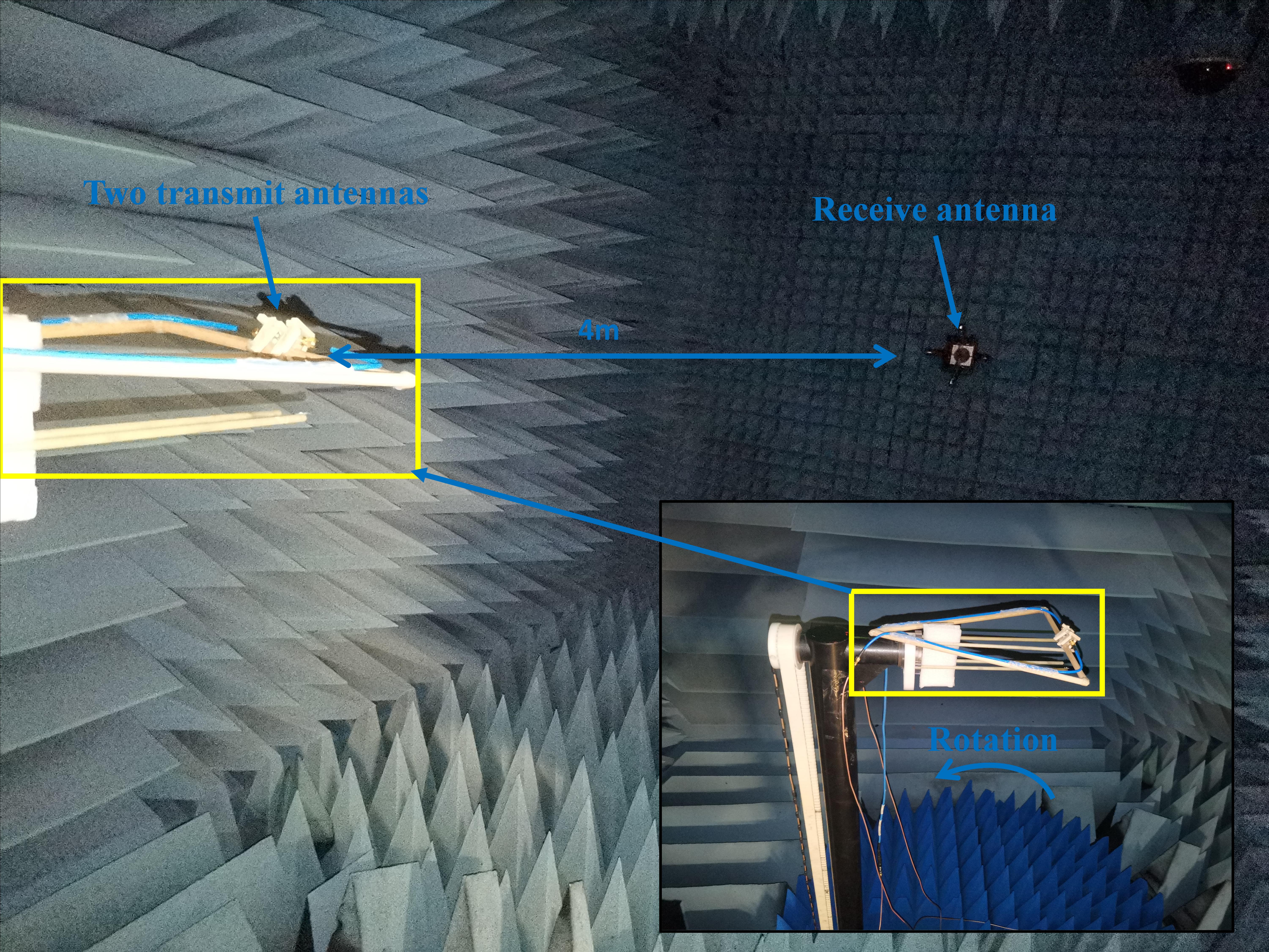}
	\caption{A photo of the experimental system in the anechoic chamber.}
	\label{fig_5S}
\end{figure}

The power measurements are carried out in a microwave anechoic chamber (see Fig. \ref{fig_5S}), which has dimensions of $7.4$m in length, $3.75$m in width and $3.75$m in height. The shielding effectiveness of EM waves with vertical and horizontal polarization are approximately $111.1$ dB and $111.5$ dB, respectively. The temperature and humidity are controlled within the ranges of 22$^\circ C$ to 24 $^\circ C$ and 41$\%$ to 45$\%$, respectively.  

\begin{figure}[htbp] 
	\centering
	\includegraphics[width=0.45\linewidth]{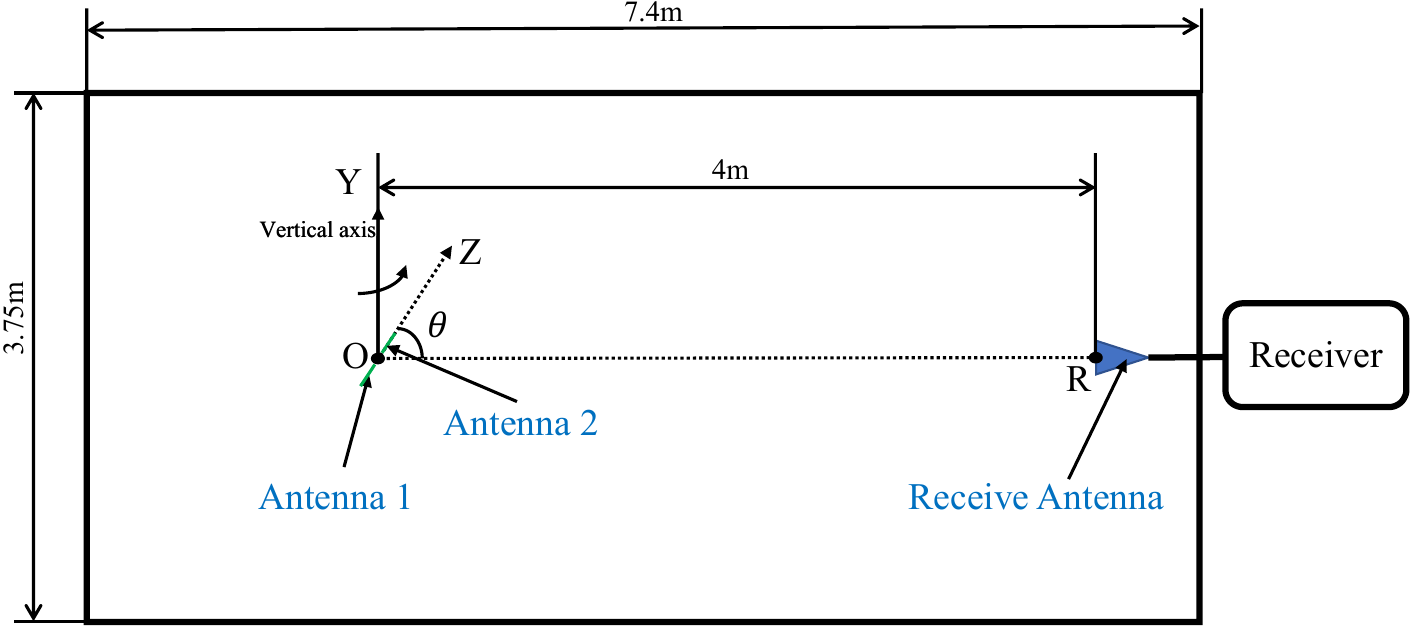}
	\caption{The measurement structure for recording the radiative powers.}
	\label{fig_4}
\end{figure}

Fig. \ref{fig_4} illustrates the geometric setup of the system for measuring the power patterns. The two antennas are aligned along the $z$-axis, with the symmetric point located at the origin ``O".  One network analyzer is used to provide co-phase EM excitations to the two antennas, while another network analyzer is connected to the receive antenna, positioned at``R", 4 meters away from the origin, to record the power data. These data are then used to plot the power patterns, of which the data are applied to Poynting’s integrals to calculate the total radiative powers. 

To obtain the power patterns over the closed spherical surface, the two antennas aligned along z-axis are rotated around the $y$-axis in the plane constructed by the $z$-axis and and the line``OR". The rotation is performed from $\theta = 0^o$ to $180^o$, with a measurement angle resolution of $1^o$.  This operation is sufficient due to the rotational symmetry with respect to the $z$-axis.  Some preparations of the experiments are carried out in the following procedures.

\subsection{Adjusting Spatial Symmetry Mode}
\begin{figure}[htbp] 
	\centering
	\includegraphics[width=0.45\linewidth]{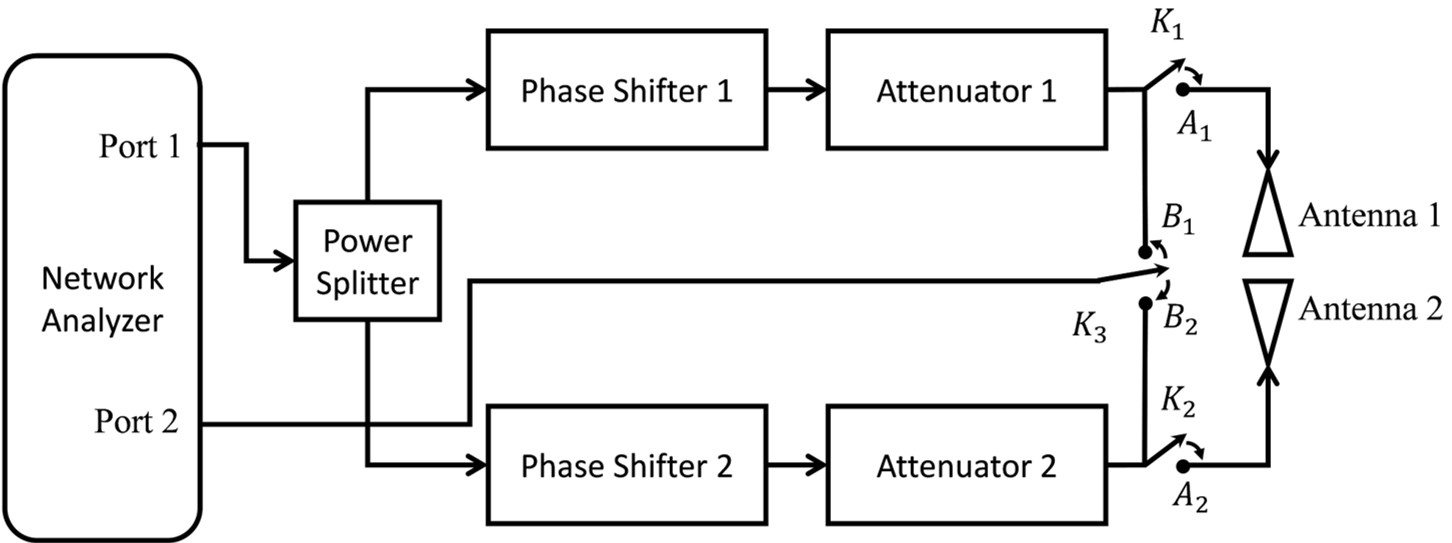}
	\caption{The circuit for adjusting the point symmetry mode.}
	\label{fig_6S}
\end{figure}

Before measuring the power patterns, we need to calibrate the radiation phases and amplitudes of the excitations of the two antennas to realize the spatial symmetric radiation-mode, using the circuit as shown in Fig. \ref{fig_6S} through the following operations.

By opening switches $K_1$ and $K_2$ and closing $K_3$ to $B_1$, the Network Analyzer provides the sinusoidal signal from port 1 to port 2.  The signal is adjusted using Phase Shifter 1 and Attenuator 1 to provide a reference signal. Next, with $K_1$ and $K_2$ still opened and $K_3$ is closed to $B_2$.  By adjusting Phase Shifter 2 and Attenuator 2 until the phase of the received signal at port 2 is $0^o$ with respect to and the power is same as those of the reference,  the construction of the symmetric radiation mode is completed.  

Finally, by opening $K_3$ from $B_1$ and $B_2$, we can perform the joint radiation of the two antennas by closing switches $K_1$ to $A_1$ and $K_2$ to $A_2$ simultaneously.  


\subsection{Individual Radiative Powers}
To obtain the individual radiative power including the coupling effect, as described in equation (10), we work as follows based on the microwave network theory [13], as follows. 

The radiative power, as described by the first equality of (10), can be calculated by 
\begin{equation}\label{13-05}
	P_i = P'_i + P''_i \quad \text{for} \ \ \ i,j = 1,2
\end{equation}
with $P'_i =Re\{Z_{ii}\}I^2_i$ and $P''_i=Re\{Z_{ij}\}I_iI_j$, 
where $P_i$ is the radiative power in present of coupling effect, $P'_i$ is the radiative power from antenna $i$ without radiation-participation of antenna $j$,  $P''_i$ is the power due to radiation from antenna $j$, i.e., the power due to the EM coupling, $Z_{ii}$ and $Z_{ij}$ are the complex value of the radiation impedances of antenna $i$ and that of the coupling from antenna $j$, respectively.  

It is noted that $P'_i$ is measured by switching off the power of antenna $j$ via closing $K_i$ to $A_i$ and opining $K_j$ from $A_j$ for $i \ne j $, separately. 

Then, (17) can be expressed by    
\begin{equation}\label{13-05}
	P_i = P'_i +\eta_{ij}P'_j =  P'_i (1+ \eta_{ij}) \quad \text{for} \ \ \ i,j = 1,2
\end{equation}
where $\eta_{ij}$ is the EM coupling factor and the second equality holds because of the radiation symmetry.   

The coupling factor can be calculated by
\begin{equation}\label{13-05}
	\eta_{ij} = \frac{Re\{Z_{ij}\}}{Re\{Z_{ii}\}}
\end{equation}
where $Z{ij}$ is the value of impedance defined in [13].
\begin{table*}[htbp] 
	\linespread{1.4}\selectfont
	\caption{The experimental data and the calculated impedances.}
	\centering
	\scalebox{1}[1]{
		\begin{tabular}{|ccccc|}
			\hline
			\multicolumn{1}{|c|}{}&
			\multicolumn{1}{c|}{Voltage gains $S_{ij}$} &\multicolumn{2}{c|}{Calculated Impedances $Z_{ij}(\Omega)$}  &\multicolumn{1}{c|}{Coupling factor $\eta_{ij}$}                                                    
			\\ \hline
			\multicolumn{1}{|c|}{$(i,j)=(1,1)$}&
			\multicolumn{1}{|c|}{$0.630*e^{-122.0^{\circ}}$}&
			\multicolumn{2}{|c|}{$14.6-25.9j$}&
			\multicolumn{1}{|c|}{\multirow{2}{*}{$\eta_{12}=0.1068$}}
			\\
			\cline{1-4}
			\multicolumn{1}{|c|}{$(i,j)=(1,2)$}&
			\multicolumn{1}{|c|}{$0.040*e^{8.0^{\circ}}$}&
			\multicolumn{2}{|c|}{$1.56-1.15j$}&
			\multicolumn{1}{|c|}{\multirow{1}{*}{}}\\
			\hline
			\multicolumn{1}{|c|}{$(i,j)=(2,1)$}&
			\multicolumn{1}{|c|}{$0.040*e^{9.3^{\circ}}$}&
			\multicolumn{2}{|c|}{$1.59-1.12j$}&
			\multicolumn{1}{|c|}{\multirow{2}{*}{$\eta_{21}=0.1089$}}
			\\
			\cline{1-4}
			\multicolumn{1}{|c|}{$(i,j)=(2,2)$}&
			\multicolumn{1}{|c|}{$0.635*e^{-120.2^{\circ}}$}&
			\multicolumn{2}{|c|}{$14.6-26.9j$}&
			\multicolumn{1}{|c|}{\multirow{1}{*}{}}\\
			\hline
	\end{tabular}}
\end{table*}

The values of the impedances mentioned above can be calculated using the experimental data of the input port voltage reflection coefficient and the reverse voltage gain measured by a Network Analyzer at the two antennas as shown in 7, before measuring the power pattern, as formulated by     

\begin{equation}
	\begin{aligned}
		\left[ \begin{matrix}
			Z_{11} & Z_{12}  \\
			Z_{21} & Z_{22} \\
		\end{matrix} \right]=&{{Z}_{0}}\left( \left[ \begin{matrix}
			1 & 0  \\
			0 & 1  \\
		\end{matrix} \right]+\left[ 
		\begin{matrix}
			{{S}_{11}} & {{S}_{12}}  \\
			{{S}_{21}} & {{S}_{22}}  \\
		\end{matrix} \right] \right)
		\times {{\left( \left[ \begin{matrix}
					1 & 0  \\
					0 & 1  \\
				\end{matrix} \right]-\left[ \begin{matrix}
					{{S}_{11}} & {{S}_{12}}  \\
					{{S}_{21}} & {{S}_{22}}  \\
				\end{matrix} \right] \right)}^{-1}}
	\end{aligned}
\end{equation}
where $Z_0=50 \Omega$ is the characteristic impedance, $S_{ij}$ is the input port voltage reflection coefficient and the reverse voltage gain for $i=j$ and $i \ne j$, respectively. 

\begin{figure}[htbp] 
	\centering 
	\includegraphics[width=0.45\linewidth]{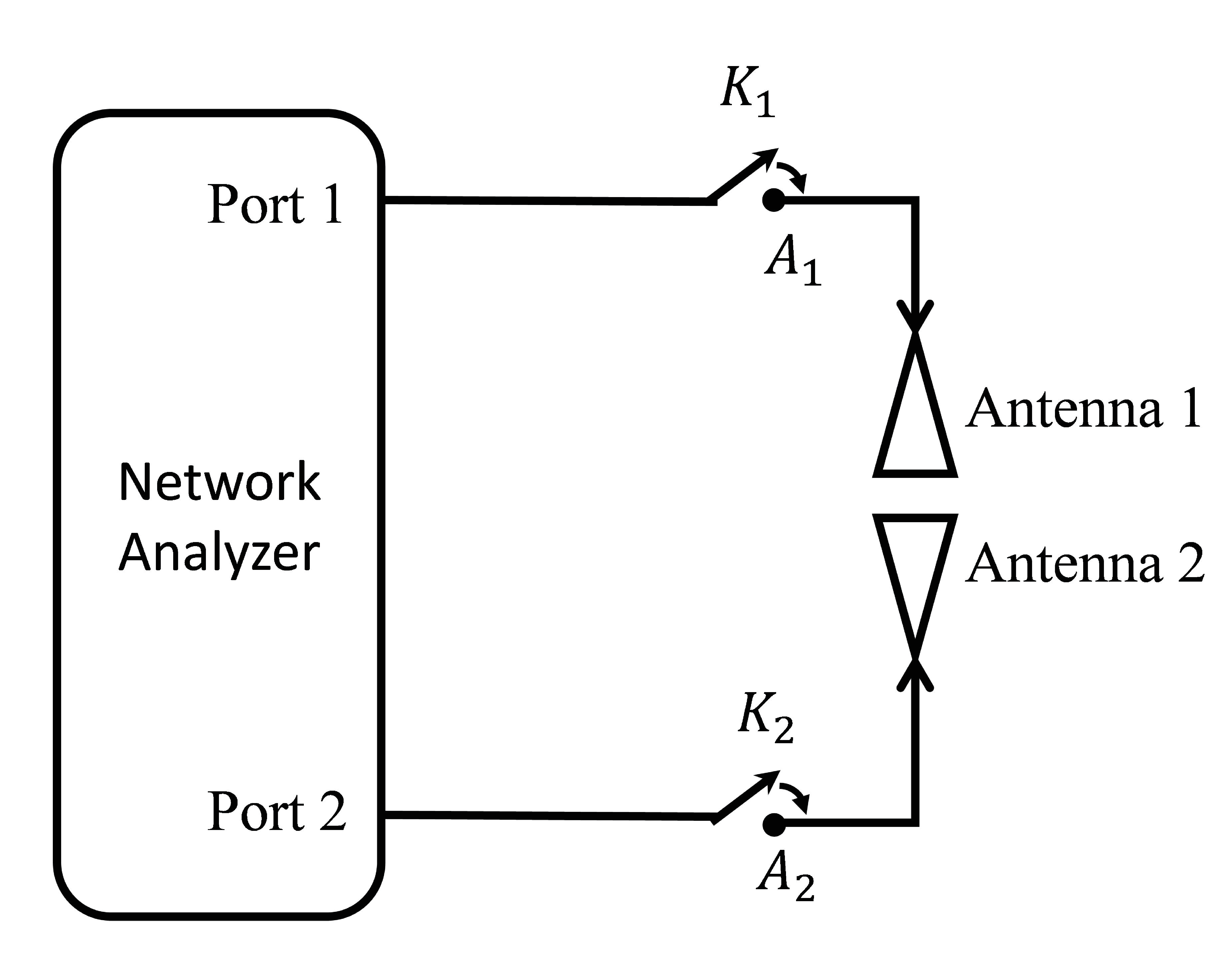}
	\caption{The circuit for measuring $S_{ij}$}
	\label{fig_7S}
\end{figure}

The data of $S_{ij}$ are measured experimentally using the circuit shown in Fig. \ref{fig_7S} as follows. By using Port $i$ as the transmitter and closing $K_i$ while leaving $K_j$ open, the value of $S_{ii}$ can be measured at Port $i$ for $i=1,2$.  While, using port $i$ as the transmitter and closing $K_1$ and $K_2$ simultaneously, $S_{ij}$ is measured at Port $j$.  Both the experimental and calculated results are listed in Table II.  In principle,  $\eta_{12}= \eta_{21}$ is expected due to the symmetry of the system.  It is noted that a negligible difference is found due to practical imperfections.  

Very important again as mentioned previously, we can found that the EM coupling between the two antennas are very small by noting the value of $\eta_{12}$ or $\eta_{21}$ in Table II, where the values are an order of magnitude small compared to 1. This indicates that our experiments adheres approximately to the theoretical zero coupling.  In this case, the experimental confirmation of the energy-doubling effect will indicate that \textbf{using the interactions between the two sources is wrong in the explanation of [1]}.

\begin{figure}[htbp] 
	\centering
	\includegraphics[width=0.45\linewidth]{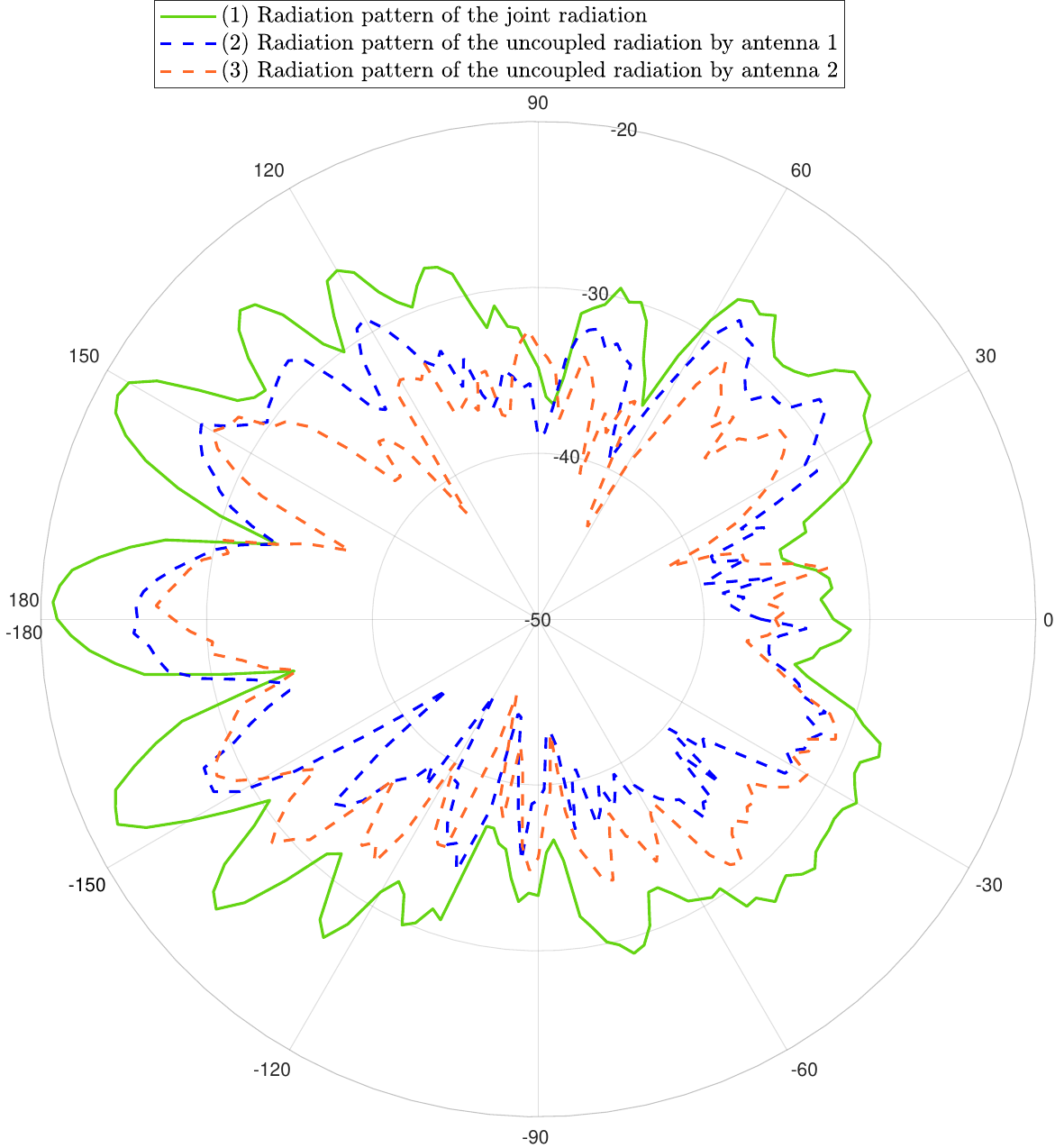}
	\caption{ Results of the radiation patterns of the joint radiation and uncoupled radiation by antenna 1 and 2, expressed in decibel.}	
	\label{fig_8}
\end{figure}

\subsection{Experimental Result}
\begin{table}[ht] 
	\linespread{1.3}\selectfont
	\caption{The results of  Radiative powers and EAF }
	\centering
	\scalebox{1}[0.9]{
		\begin{tabular}{|ccccccccc|}
			\hline
			\multicolumn{9}{|c|}{The radiative powers }                                                                                                                          \\ \hline
			\multicolumn{3}{|c|}{Antenna 1} & 
			\multicolumn{3}{|c|}{Antenna 2} & 
			\multicolumn{3}{|c|}{Joint radiation}                                                   \\ \cline{1-9} 
			\multicolumn{3}{|c|}{$P_1 \approx P'_1$} &      
			\multicolumn{3}{|c|}{$P_2 \approx P'_2$} &
			\multicolumn{3}{|c|}{$P_T$}  \\ \hline
			\multicolumn{3}{|c|}{0.0092mW} &         
			\multicolumn{3}{|c|}{0.0073mW} &
			\multicolumn{3}{|c|}{0.0263mW}
			\\ \hline
	\end{tabular}}
\end{table}

The experimental verification of the energy-doubling effect is carried out as follows. 

After adjusting the symmetric mode, the total radiative power pattern is measured by activating both antennas simultaneously, following the scheme described above.  The power pattern is obtained as shown in Fig.8.  Taking the data of the power pattern to the Poynting integral over the close surface, the total radiation power of the superposed wave,calculated using (14) is shown in Table III.      

The individual radiative power pattern of antenna $i$ is measured by switching on the power of antenna $i$ and off the power of antenna $j$ for $i \ne j$.  The result is also shown in Fig 8, which provides the data to be taken to the integral over the entire surface.  Then, $P'_i$ is obtained for $i=1,2$.  Further, by neglecting the coupling factor (as mentioned above), $P_i$ described in (10) or and (18) is obtained as shown in Table III.   

The experimental result of (16) is finally obtained as   
\begin{equation}
	\begin{split}\label{03-1}
		\beta = \frac{P_T}{P_1+P_2} = 1.59 
	\end{split}
\end{equation}  
which substantiates the conclusion of the energy-doubling effect with $\beta =2$ in ideal dipole model.  

Actually, the discrepancy of $\beta$ from $1.59$ to $2$ (more precisely to $1.96$ obtained by the simulation as shown in Fig.2) can be attributed to geometric deviations from the idealized configuration of the two simulated antennas. This is due to the nonzero cross-sectional areas of the antennas, as well as the limited accuracy in the measurements, e.g., result in $P'_1 \ne P'_2$.   

Nonetheless, all the results in this work agree with the energy-doubling effect and the fact that the energy conservation law does not necessarily hold with the superposed wave.   

\section{Conclusion}
Upon spatial symmetry the superposition law is applied to the waves from the co-phase sources, revealing the \textbf{existence of the energy-doubling effect}—a phenomenon that contradicts the energy conservation law. The theoretical analysis leads to this key conclusion, and the experimental result verifies it with an observed increase of 1.59 times. Furthermore, the simulation results indicate \textbf{an energy fluctuation that ultimately converges to the value required by the law of energy conservation}.  This means the law of energy conservation dominates at large distances between the two sources, such as when the distance is several times greater than the wavelength. In addition, the mistake in the explanation of [1] is revealed theoretically and experimentally.  Our findings may provide some insights for further research into the energy behavior of EM wave superposition.

%
%

\section*{acknowledgments}
The experiments are taken in the microwave anechoic chamber of TsingHua University. 









\end{document}